\documentclass[a4paper,3p,times]{elsarticle}
\usepackage[utf8]{inputenc}
\usepackage{amsmath}
\usepackage{amsfonts}
\usepackage{graphicx}
\usepackage{float}
\usepackage{bm}
\usepackage{hyperref}
\usepackage{color}
\usepackage{amssymb}
\usepackage{dsfont}
\usepackage{slashed}
\usepackage{hyperref}
\hypersetup{
    colorlinks=true,
    urlcolor=blue,
    linkcolor = blue,
    urlcolor  = blue,
    citecolor = blue,
    anchorcolor = blue,
    pdftitle={SU3-Lambda},
    pdfsubject={SU3-Lambda}}
\allowdisplaybreaks
\makeatletter
\def\ps@pprintTitle{%
	\let\@oddhead\@empty
	\let\@evenhead\@empty
	\def\@oddfoot{}%
	\let\@evenfoot\@oddfoot}
\makeatother

\begin{document}
\title{New insights into the nature of the $\Lambda(1380)$ and $\Lambda(1405)$ resonances away from the SU(3) limit}
\author[itp,ucas,peng]{Feng-Kun Guo}
\author[bonn]{Yuki Kamiya}
\author[bonn,gwu]{Maxim Mai}
\author[bonn,fzj,tbilisi]{Ulf-G. Mei\ss{}ner}
\address[itp]{CAS Key Laboratory of Theoretical Physics, Institute of Theoretical Physics,\\
	Chinese Academy of Sciences, Beijing 100190, China}
\address[ucas]{School of Physical Sciences, University of Chinese Academy of Sciences, Beijing 100049, China}
\address[peng]{Peng Huanwu Collaborative Center for Research and Education, Beihang University, Beijing 100191, China}
\address[bonn]{Helmholtz-Institut f\"ur Strahlen- und
	Kernphysik and Bethe Center for Theoretical Physics,\\
	Universit\"at Bonn, D-53115 Bonn, Germany}
\address[fzj]{Institute for Advanced Simulation, Institut f\"ur Kernphysik and
	J\"ulich Center for Hadron Physics,\\ Forschungszentrum J\"ulich,
	D-52425 J\"ulich, Germany}
\address[tbilisi]{Tbilisi State University, 0186 Tbilisi, Georgia}
\address[gwu]{Institute for Nuclear Studies and Department of Physics, The George Washington University, Washington, DC 20052, USA}
\begin{abstract}
 Starting from the SU(3) limit, we consider the nature of the dynamically generated resonances $\Lambda(1380)$, $\Lambda(1405)$ and $\Lambda(1680)$ as the pion and kaon
 masses are tuned to their physical values. We show that the accidental symmetry of the
 two octets due to the leading order Weinberg-Tomozawa term is broken by the next-to-leading order terms. Most interestingly, we observe an interchange of the two trajectories 
 of the $\Lambda(1380)$ and the $\Lambda(1405)$ away from the SU(3) limit at next-to-leading order. This remarkable phenomenon can be investigated using lattice QCD calculations that start from the SU(3) limit.
\end{abstract}
\maketitle

\section{Introduction}

The $\Lambda$(1405) is a truly exotic hadron state especially after in the framework of unitarized chiral perturbation
theory its two-pole structure was revealed and understood~\cite{Oller:2000fj,Jido:2003cb}. These two states
are now called $\Lambda$(1380) and  $\Lambda$(1405) \cite{ParticleDataGroup:2022pth}. For recent reviews, see Refs.~\cite{Mai:2020ltx,Hyodo:2020czb,Meissner:2020khl}, and a state-of-the-art calculation with the chiral potential
expanded to the next-to-next-to-leading order is given in Ref.~\cite{Lu:2022hwm}. Also, an uncertainty analysis supplemented
with an investigation of the correlations between the various poles is prsented in Ref.~\cite{Sadasivan:2022srs}.

Lattice studies of the $\Lambda(1405)$ 
are not very abundant, for earlier works see,  e.g.,~\cite{Engel:2012qp,Hall:2014uca}. Of particular interest is the very recent coupled-channel work \cite{Bulava:2023rmn,Bulava:2023gfx} that also reported two poles consistent with the phenomenological values for $M_\pi \simeq 200$~MeV and $M_K\simeq 487$~MeV.  In that study, the lower pole is a virtual state, whereas the higher pole is a resonance.

It is also known that the SU(3) limit, where the $u$, $d$, and $s$ quarks 
have the same  mass, can be favorably used in lattice QCD, see,  e.g.,~\cite{Bietenholz:2011qq}.
To assist such lattice studies and also to gain deeper insight, we
study here in detail the development of the poles from the SU(3) flavor limit to the physical masses of the Goldstone bosons. 

To be more precise, in the SU(3) limit the interaction of the Nambu-Goldstone (NG) mesons and the octet baryons can be decomposed into group irreducible representations (irreps). By setting up the quark configurations so that they are in the
appropriate group representation,  we can investigate the interaction in each representation.
It was already pointed out that the origin of the two-pole  structure of the $\Lambda(1405)$ and $\Lambda(1380)$ resonances is related to the strength of the attractive forces of the decomposed interaction~\cite{Jido:2003cb}.
It was found there that the $\Lambda(1405)$ ($\Lambda(1380)$) pole is connected to the attractive interaction originated in octet (singlet)
using a simple extrapolation to the SU(3) limit. 
Thus, checking these interaction components in the SU(3) limit is helpful in understanding the nature of the $\Lambda(1380), \Lambda(1405)$ and also checking the chiral dynamics. 
However, such results on the extrapolation from the physical point to the SU(3) limit or some other limit can depend on the detailed scheme of the extrapolation.
As found in Ref.~\cite{Nagahiro:2011jn}, where the $a_1(1260)$ is studied varying the number of colors, the trajectory of the two poles in that case may flip depending on the model details. 
Thus, to obtain a concrete picture for a given resonance, a detailed scheme for the extrapolation where  reasonable parameters are chosen at any hadron mass point in the extrapolation is required. 

In this work, we revisit the origin of the $\Lambda(1405)$ and $\Lambda(1380)$ states within chiral dynamics by considering the detailed extrapolation to the SU(3) limit,  employing the results from chiral perturbation theory that give the the quark-mass dependence of the
GB and baryon masses of relevance here.  We further consider the $\Lambda(1680)$
that is also generated from the octet. See also Refs.~\cite{Garcia-Recio:2003ejq,Molina:2015uqp} for earlier work along these lines, and  Refs.~\cite{Bruns:2021krp,Xie:2023cej} for recent ones. In our work, we go beyond the Weinberg-Tomozawa (WT) term  approximation for the meson-baryon interaction potential, and, in particular, we include the next-to-leading order (NLO) operators.

This article is organized as follows. In Sect.~\ref{sect:formalism}, we summarize our approach and the scheme for the extrapolation. In Sect.~\ref{sect:extra_with_masses}, we explain how our model parameters are determined. In Sect.~\ref{sect:extra_with_masses}  we collect all ingredients of our extrapolation formulas and discuss the
evolution of the two poles from the SU(3) limit as the quark masses are tuned to their physical values.
Finally, we summarize our study in Sect.~\ref{sect:summary}. In~\ref{app}, we investigate the poles with quark masses used in the recent lattice calculation~\cite{Bulava:2023rmn,Bulava:2023gfx}.

\section{Formalism}
\label{sect:formalism}

\subsection{Chiral unitary amplitude}

In the $\bar{K}N$ system, the WT term, which appears at leading order (LO) in the chiral expansion, gives the dominant contribution to the $S$-wave interaction. At the same order two one-baryon exchange diagrams contribute, the so-called $s$- and $u$-channel Born terms. As common in many studies, we do not consider such LO Born terms here which are negligible in the $S$-wave projected amplitudes at low energies. The WT interaction term projected to the $S$-wave and isospin $I=0$ is given by\footnote{ We have updated the notation compared to the journal version. No calculation or results are changed by this.} 
\begin{align}
V_{ij}^{\mathrm{WT}}(\sqrt{s}) = -\frac{C_{ij}^{\mathrm{WT}}}{4F^2} \mathcal{N}_i\mathcal{N}_j (2\sqrt{s}-m_i-m_j)
 ~~~~\text{for}~~~~
    i,j\in\{\pi\Sigma,\bar KN,\eta\Lambda,K\Xi\},
\label{eq:V}
\end{align}
with the meson decay constant $F$, the normalization factor $\mathcal{N}_i=\sqrt{m_i+E_i}$, the baryon mass $m_i$, the baryon energy $E_i=\sqrt{m_i^2 + q_i^2}$, and $q_i=\sqrt{(s-(M_i+m_i)^2)(s-(M_i-m_i)^2)}/(2\sqrt{s})$. The coefficients $C_{ij}^{\mathrm{WT}}$ are calculated from the LO chiral Lagrangian and are provided explicitly in, e.g., Refs.~\cite{Oset:1997it,Borasoy:2005ie}. 

In the SU(3) limit, this interaction term can further be decomposed into irreps. For the meson-baryon system from the octet of the Nambu-Goldstone (NG) bosons and the octet of ground state baryons, the corresponding decomposition of multiplets reads  $\{1, 8_{\rm s},8_{\rm a}, 10, \overline{10}, 27\}$ with the subscripts ``s'' and ``a'' referring  to the symmetric and antisymmetric representations, respectively, see for more details Refs.~\cite{Bruns:2021krp,Hyodo:2006kg}. Projecting the above two-particle isoscalar states to the relevant multiplets is accomplished by 
\begin{align}
	\left(
	\begin{array}{l}
		|\pi\Sigma\rangle\\
		|\bar{K}N\rangle\\
		|\eta\Lambda\rangle\\
		|K\Xi\rangle
	\end{array}\right) 
	=\frac{1}{\sqrt{40}}
	\left(
	\begin{array}{cccc}
		\sqrt{15} & -\sqrt{24} & 0 & -1 \\
		-\sqrt{10} & -2            & \sqrt{20} & -\sqrt{6} \\
		-\sqrt{5} & -\sqrt{8} & 0 & 3\sqrt{3} \\
		\sqrt{10} & 2 & 2\sqrt{5} & \sqrt{6}
	\end{array} \right)
	\left(
	\begin{array}{l}
		|1\rangle\\
		|8\rangle\\
		|8^\prime\rangle\\
		|27\rangle
	\end{array}\right)\,,
\end{align}
taken from Ref.~\cite{Roca:2008kr} but correcting for the $i=j=4$ entry, where $|8\rangle$ and $|8'\rangle$ are mixtures of the symmetric and antisymmetric octets. With this transformation, the WT term diagonalizes as 
\begin{align}
    C_{\alpha\beta} = \left(
	\begin{array}{cccc}
		6 & 0 & 0 & 0 \\
		0 & 3 & 0 & 0 \\
		0 & 0 & 3 & 0 \\
		0 & 0 & 0 & -2 \\
	\end{array} \right)
~~~~\text{for}~~~~
    \alpha,\beta\in\{1,8,8^\prime,27\} .
\label{eq:Valpha}
\end{align}
The crucial observation at this point is that the singlet ($1$) representation gives the strongest attractive interaction (positive coefficient)  while the octet ($8$) representations also give attraction, noting that the degeneracy of the octets is an accidental symmetry of the WT term, as already discussed in~\cite{Bruns:2021krp}. Noteworthy, the symmetry is broken 
already by the LO Born terms, but also 
by the NLO contact terms discussed below. The (27) representation shows a repulsive (negative coefficient) interaction. The ($10,\overline{10}$) irreps have zero interaction at LO and are
not further considered. 

We also include the NLO terms, which split into the so-called symmetry breakers ($C^{\rm NLO1}(b_0,b_d,b_F)$) and dynamical terms ($C^{\rm NLO2}(d_1,d_2,d_3,d_4)$). We note that the former amplitude vanishes in the chiral limit, but the latter does not. Overall, the NLO potential projected to the $S$-wave and $I=0$ is given by
\begin{align}
	V_{ij}^{\rm NLO}(\sqrt{s}) = 2\frac{\mathcal{N}_i\mathcal{N}_j}{F^2}
    \left(
        C_{ij}^{\rm NLO1}-2C_{ij}^{\rm NLO2}
            \left(
            E_iE_j+\frac{q_i^2 q_j^2}{3\mathcal{N}_i^2\mathcal{N}_j^2}
            \right)
    \right)
 ~~~~\text{for}~~~~
    i,j\in\{\pi\Sigma,\bar KN,\eta\Lambda,K\Xi\}.
\label{eq:VNLO}
\end{align}
The coefficients $C^{\rm NLO1}$ and $C^{\rm NLO2}$ are read off from the NLO chiral Lagrangian, 
which are divided by two from those found in Ref.~\cite{Borasoy:2005ie} to be consistent with the notation in this work.
Projected to the relevant multiplets these coefficients read
\begin{align}
    C^{\rm NLO1}_{\alpha\beta} &= \left(
	\begin{array}{cccc}
		\frac{4}{3}(3b_0+7b_D)m_q & 0 & 0 & 0 \\
		0 & \frac{2}{3}(6b_0+b_D)m_q & -\sqrt{20}b_Fm_q & 0 \\
		0 & -\sqrt{20}b_Fm_q & 2(2b_0+3b_D)m_q & 0 \\
		0 & 0 & 0 & 4(b_0+b_D)m_q \\
	\end{array} \right),
\\
    C^{\rm NLO2}_{\alpha\beta} &= \left(
	\begin{array}{cccc}
		-3d_2+\frac{9}{2}d_3+d_4& 0 & 0 & 0 \\
		0 & \frac{1}{2}(-3d_2+d_3+2d_4)& -\frac{\sqrt{5}}{2}d_1 & 0 \\
		0 & -\frac{\sqrt{5}}{2}d_1 & \frac{1}{2}(9d_2-d_3+2d_4) & 0 \\
		0 & 0 & 0 & \frac{1}{2}(2d_2+d_3+2d_4) \\
	\end{array} \right).
\label{eq:Valpha}
\end{align}
These NLO terms lift the accidental symmetry of the two octets in the SU(3) limit. 
Note that the Born terms, which we neglect due to its small contribution to the $S$-wave, also breaks this symmetry. 
The low-energy constants (LECs) of the 
symmetry breakers 
($b_0$, $b_D$, and $b_F$)
will be determined from the baryon masses as discussed below and the dynamical LECs $(d_1$, $d_2$, $d_3$, and $d_4)$ from a fit to the cross sections $K^-p\to K^-p, \bar{K}^0n, \pi^0\Lambda, \pi^0\Sigma^0, \pi^\mp \Sigma^\mp$ and usual threshold ratios. For details of these data including data repositories, see the recent review~\cite{Mai:2020ltx}.

The attractive interaction between the meson-baryon pair leads to a dynamical generation of the $\Lambda(1405)$ and $\Lambda(1380)$ states. For this a Bethe-Salpeter equation is realized using either WT or WT plus NLO potential in a matrix equation
for the unitary scattering amplitude $T$,
\begin{align}
\label{eq:BSE}
	T_{ij} = V_{ij} + \sum_k V_{ik}G_kT_{kj}\,
~~~~{\rm for} \quad 
    i,j,k\in\{\pi\Sigma,\bar KN,\eta\Lambda,K\Xi\},
\end{align}
with $V$ the potential matrix.
For the inclusion of off-shell terms and higher partial waves in this type of dynamical equations, see Refs.~\cite{Bruns:2010sv,Mai:2012dt,Sadasivan:2018jig}. Here, the loop function is defined as 
\begin{align}
G_i(\sqrt{s}) = i\int\frac{d^4q}{(2\pi)^4} \frac{1}{(P-q_i)^2-m_i^2 + i\epsilon}\frac{1}{q_i^2-M_i^2+i\epsilon}, 
\end{align}
where $M_i$ ($q_i$) is the meson mass (momentum) of channel $i$, and the total momentum $P$ is given by $P^\mu = (\sqrt{s},\bm{0})$. Using dimensional regularization, this loop function is given as~\cite{Oller:2000fj} 
\begin{align}
\label{eq:G_loop}
	G_i(\sqrt{s}) = \frac{1}{16\pi^2} \bigg\{&a_i(\mu) +\ln\frac{m_i^2}{\mu^2} + \frac{M_i^2-m_i^2+s}{2s}\ln\frac{M_i^2}{m_i^2} 
	+\frac{\bar{q}_i}{\sqrt{s}}  
	\Big[\ln\left(s - (M_i^2 - m_i^2) + 2\bar{q}_i\sqrt{s}\right) \\ 
	&+\ln\left(s + (M_i^2 - m_i^2) + 2\bar{q}_i \sqrt{s}\right)  
	-\ln\left(-s + (M_i^2 - m_i^2) + 2\bar{q}_i \sqrt{s}\right) 
	-\ln\left(-s - (M_i^2 - m_i^2) + 2\bar{q}_i\sqrt{s}\right)   \Big]\bigg\}, \nonumber 
\end{align}
where $\bar{q}_i$ is the magnitude of three-momentum in the center-of-mass (c.m.) frame, $\mu$ is the scale of dimensional regularization, and $a_i(\mu)$ is the channel-dependent subtraction constant. The values of the subtraction constants in Eq.~\eqref{eq:G_loop} control the short-distance contributions that are not explicitly considered in the model. It is known that, if the dynamical nature of the system under consideration is well controlled by the chiral interaction, the natural value of $a_i(\mu=1~{\rm GeV})$ should be around $-2$ at the physical point, see Ref.~\cite{Oller:2000fj}. In phenomenological studies, their values are usually fixed according to a given procedure (see below) or tuned so that the experimental data or lattice results are well reproduced. However, the subtraction constant depends on details of the system, e.g., hadron masses. Thus, these values can  gradually change in the extrapolation from the physical point to the SU(3) limit. There are different procedures of fixing the subtraction constants. Here, we follow the scheme proposed in Refs.~\cite{Lutz:2001yb,Hyodo:2008xr}, which requires the loop function to satisfy the relation
\begin{align}
	G(\sqrt{s} = m_B;a(\mu)) = 0 ~~~\Longleftrightarrow~~~ T(\mu)=V(\mu)\,,
 \label{eq:natural_subt}
\end{align}
with $\mu=m_B$ the physical mass of the involved baryon. In this way, given the hadron masses, we can determine a value of the subtraction constant. Note that, in the SU(3) limit, all the baryon masses are equal so that there is effectively only a single subtraction constant.

\subsection{Quark mass dependence of the NG bosons}

Up to one-loop order, the pion mass dependence of the NG boson masses $\{M_P|P=\pi, K, \eta\}$ are provided through chiral perturbation theory~\cite{Gasser:1984gg} reading
%
\begin{align}
	M_\pi^2=&\, M_{0\,\pi}^2\left[1+\mu_\pi-\frac{\mu_\eta}{3}+\frac{16
		M_{0\,K}^2}{F_0^2}\left(2L_6^r-L_4^r\right) + 
	\frac{8M_{0\,\pi}^2}{F_0^2}\left(2L_6^r+2L_8^r-L_4^r-L_5^r\right)
	\right]\,, \nonumber\\
	M^2_K=&\, M^2_{0\,K}\left[1+\frac{2\mu_\eta}{3}+\frac{8
		M_{0\,\pi}^2}{F_0^2}\left(2L_6^r-L_4^r\right) + \frac{8
		M_{0\,K}^2}{F_0^2}\left(4L_6^r+2L_8^r-2L_4^r-L_5^r\right)\right]\,,\nonumber\\
	M^2_\eta=&\, M^2_{0\,\eta} \left[1+2\mu_K-\frac{4}{3}\mu_\eta+
	\frac{8M^2_{0\,\eta}}{F_0^2}(2L_8^r-L_5^r)+
	\frac{8}{F_0^2}(2 M^2_{0\,K}+M^2_{0\,\pi})(2L_6^r-L_4^r)
	\right] \label{eq:mass_meson} \\ 
	&+M^2_{0\,\pi}\left(-\mu_\pi+\frac{2}{3}\mu_K+\frac{1}{3}\mu_\eta\right)
	+\frac{128}{9F_0^2}(M^2_{0\,K}-M^2_{0\,\pi})^2(3L_7^{}+L_8^r)\,,
	\nonumber 
\end{align}
where $M_{0P}$ is the NG boson mass at leading chiral order, and $L_i^{(r)}$ are the renormalized NLO LECs. Furthermore, $\mu_P=M_{0\, P}^2/(32 \pi^2 F_0^2)\log(M_{0 \,P}^2/\mu^2)$ for $P=\pi,K,\eta$. Note that the LEC $L_7$ does not get renormalized and that in Ref.~\cite{Gasser:1984gg} the LO masses inside the square brackets are substituted by the physical masses. The latter is allowed as the difference from the above expressions is of higher (two-loop) order. The LO masses are related to the constant $B_0$ and the quark masses $\hat{m} =(m_u+m_d)/2$ and $m_s$ as
\begin{align}
	M_{0\,\pi}^2= 2 \hat m B_0\,, \qquad 
	M_{0\,K}^2=(\hat m + m_s) B_0\,, \qquad
	M^2_{0\,\eta} = {2\over3}(\hat m+2m_s)B_0\,,
	\label{eq:LOmasses}
\end{align}
where isospin breaking effects have been neglected. The relations for the decay constants of the NG bosons are given as~\cite{Gasser:1984gg} 
\begin{align}
	F_\pi=& F_0\left[1-2\mu_\pi-\mu_K+\frac{4
		M_{0\,\pi}^2}{F_0^2}\left(L_4^r+L_5^r\right)+\frac{8
		M_{0\,K}^2}{F_0^2}L_4^r\right]\,, \nonumber \\
	F_K=&
	F_0\left[1-\frac{3\mu_\pi}{4}-\frac{3\mu_K}{2}-\frac{3\mu_\eta}{4}+\frac{4
		M_{0\,\pi}^2}{F_0^2}L_4^r +\frac{4
		M_{0\,K}^2}{F_0^2}\left(2L_4^r+L_5^r\right)\right]\,, \label{eq:f_eta}
	\\
	F_\eta=&F_0\left[1-3\mu_K+ \frac{4
		L_4^r}{F_0^2}\left(M_{0\,\pi}^2+2M_{0\,K}^2\right)
	+\frac{4M_{0\,\eta}^2}{F_0^2}L_5^r\right]\,.\nonumber
\end{align}
Approaching the SU(3) limit these quark mass dependencies simplify strongly and can be written compactly as 
\begin{align}
	M_{\mathrm{SU(3)}}   &=M_{0,\mathrm{SU(3)}}^2 \left[1 - \frac{2}{3} \mu_{\mathrm{SU(3)}} \frac{8M_{0,\mathrm{SU(3)}}^2}{F_0^2}(-3L_4^r - L_5^r + 6L_6^r + 2L_8^r) \right]\,, \nonumber\\
	F _{\mathrm{SU(3)}}    &= F_0\left[1-3\mu_\mathrm{SU(3)} +\frac{4M_{0,\mathrm{SU(3)}}^2}{F_0^2}(3L_4^r +L_5^r)\right]\,, \nonumber
\end{align}
where $M_{\mathrm{SU(3)}}$ and $F _{\mathrm{SU(3)}} $ are the meson mass and decay constant in the SU(3) limit, $M_{0,\mathrm{SU(3)}}$ is the LO mass, and $F_0$ is decay constant in the chiral limit and does not depend on the quark masses. Furthermore, $\mu_{\mathrm{SU(3)}} = M_{0,\mathrm{SU(3)}}^2/(32\pi^2F_0^2)$  $\times\log(M_{0,\mathrm{SU(3)}}^2/\mu^2)$.

In principle, we may use the LO meson masses and decay constants to be fully consistent with the chiral orders considered in the baryon sector. Nevertheless, here we take the NLO ones to have a better description of the quark mass dependence and the difference is a subleading effect.
In this study, we employ the $L_i$ values obtained in Fit 1 in Ref.~\cite{Nebreda:2010wv} where the  experimental and lattice meson-meson scattering data are fitted.
In Sect.~\ref{sect:extra_with_masses}, we determine the $M_{0P}$ values by fitting to the physical and lattice meson masses.

\subsection{Quark mass dependence of octet baryon masses}

For the octet baryon masses, the chiral extrapolations are written down up to the fourth order in chiral expansion in Refs.~\cite{Borasoy:1996bx,Frink:2004ic}. Within chiral perturbation theory, the baryon mass is written down generally as 
\begin{align}
	m_B = m_0 + m_B^{(2)} + m_B^{(3)} + m_B^{(4)} + \cdots,
\end{align} 
where $m_0$ is the baryon mass in the chiral limit ($\hat m = m_s = 0$), and $m_B^{(i)}$'s are the terms of the order of $M_P^i$. Here, we take into account terms up to $m_B^{(2)}$ consistent with the treatment in Eq.~\eqref{eq:VNLO}. 
In this work, we do not go to the fourth order in the chiral expansion for the baryon properties to avoid the appearance of large kaon and eta loops and the related convergence problems~\cite{Borasoy:1996bx}.
In the  isospin symmetric case, the LO term of $m_B^{(2)}$ is given as 
\begin{align}
	m_B^{(2)} = \gamma_{1,B}B_0\hat{m} + \gamma_{2,B} B_0 m_s, \label{eq:m_baryon}
\end{align}
where the coefficients $\gamma_{i,B}$ are 
\begin{align}
	\gamma_{1,\Sigma}   & = - 8(b_0+ b_D),  \quad \gamma_{2,\Sigma}     = -4b_0, \quad
	\gamma_{1,N},  = - 8b_0 -4(b_D+b_F),   \quad  \gamma_{2,N}  = 4(-b_0 - b_D + b_F), 
 \nonumber\\
	\gamma_{1,\Xi} &= - 8b_0 +4(-b_D+b_F),  \quad  \gamma_{2,\Xi}  = -4(b_0 + b_D + b_F), \quad
	\gamma_{1,\Lambda}  = - 8b_0 -\frac{8}{3}b_D, \quad \gamma_{2,\Lambda} =  -4b_0 -\frac{16}{3} b_D,  
\end{align}
where $b_0, b_D$ and $b_F$ are the aforementioned symmetry-breaking LECs. In the SU(3) limit with $\hat{m} = m_s$, this reduces to 
\begin{align}
	m_B^{(2)}=(-12b_0 -8b_D)B_0\hat{m}\,.
\end{align}
Given $m_0$, $b_0$, $b_D$ and $b_F$, the baryon masses at any unphysical quark masses can be evaluated. However, with only the physical hadron masses,  $m_0$ and $b_0$ cannot be  fixed independently. This is because these parameters give a common contribution of $(m_0-12b_0B_0\hat{m})$ to every octet baryon mass. To separate these two LECs, one would have to consider, e.g., the pion-nucleon $\sigma$ term~\cite{Borasoy:1996bx} or baryon masses at unphysical quark masses.

\begin{table}[t]
\begin{center}
	\caption{The determined parameters for the extrapolation. For the mesons, the LO masses  and the meson decay constant in the chiral limit from the fit  are listed. For the baryons, the chiral limit mass $m_0$ and the symmetry breaking LECs are listed.
	\label{tab:params_extra}}
    \begin{tabular}{cccc|cccc}
		\hline\hline
    $M_{0,\pi}$ [MeV]& $M_{0,K}$ [MeV]&  $F_0$ [MeV]  & $L_7$ &
    $m_0$ [MeV]& $b_0$ [GeV$^{-1}$]& $b_D$ [GeV$^{-1}$] & $b_F$ [GeV$^{-1}$] \\
    \cline{1-4}\cline{5-8}
    125.0 & 443.4& 79.3 & $-4.48 \times 10^{-4}$  &
    844.9&$-0.43$ &$0.08$ & $-0.27$\\
	\hline\hline
    \end{tabular}
\end{center}
\end{table}

\begin{table}[t]
\begin{center}
\caption{Meson and baryon masses as well as meson decay constants calculated with the fitted LECs. All units are MeV.
\label{tab:mass}}
\begin{tabular}{cccc | ccc | ccc}
	\hline\hline
    $m_N$ & $m_\Lambda$& $m_\Sigma$  & $m_\Xi$ &
    $M_\pi$ & $M_K$ &$M_\eta$ &
    $F_\pi$ & $F_K$ &$F_\eta$ \\
    \cline{1-4}\cline{5-7}\cline{8-10}     
    940.9&1111.5 &1191.6&1322.2 &
    137.3& 495.6& 547.9 &
    92.4 & 112.7 & 121.7\\
	\hline\hline
\end{tabular}
\end{center}
\end{table}

\section{Nature of the $\Lambda$ resonances for varying pion and kaon masses}
\label{sect:extra_with_masses}

\subsection{Parameters and trajectories}

The quark mass dependence provided in the previous section serves as a guideline to define a trajectory from the SU(3) symmetric point to the physical one. In that, we first need to determine the pertinent LECs for the mesons and baryons. Following the procedure of Ref.~\cite{Molina:2015uqp}, $M_{0,P}$ and $F_0$ are determined by fitting to the isospin averaged values of the physical meson masses $M_\pi$, $M_K$ and $M_\eta$, and the pion decay constant $F_\pi$. In this procedure, the LEC $L_7$ is refitted so that these four physical values are well reproduced. The obtained values are summarized in Table~\ref{tab:params_extra} which then leads to the well-reproduced meson masses provided in Table~\ref{tab:mass}. In the same time the fitted $L_7$ is very close to original value of $-4.4\times 10^{-4}$~\cite{Gasser:1984gg}. 

Next, we determine the parameters $m_0, b_D$, and $b_F$ for the baryon masses.
To this end, we fit  the isospin averaged values of the octet baryon masses $m_N, m_{\Lambda}$ and $m_{\Sigma}$. Additionally, the baryon masses were obtained in the lattice QCD  calculation at unphysical quark masses in,  e.g., Ref.~\cite{Hall:2014uca}, and using these values we determine $m_0$ and $b_0$ separately. Among the lattice sets in Ref.~\cite{Hall:2014uca}, we employ the heaviest one for which the isospin averaged baryon mass is $1444.2$~MeV and the isospin averaged octet meson mass is $659.4$~MeV. In this case, the pion mass dependence is such that the LECs $m_0$ and $b_0$ can easily be separated. The resulting parameters are also listed in Table~\ref{tab:params_extra}. The obtained value for $m_0$ is consistent with the one obtained in the two-flavor expansion in chiral perturbation theory supplemented with the pion-nucleon sigma term from Roy-Steiner equations~\cite{Hoferichter:2015hva}. The difference to that value can be attributed to the strangeness sigma term. More precisely, in~\cite{Hoferichter:2015hva} a two-flavor expansion was performed, thus the contribution $\sim m_s \langle N|\bar ss|N\rangle \simeq 30\,$MeV is subsumed in $m_0$, which is not the case here; this explains the difference of the
two values.
The $\Xi$ baryon mass, which is not fitted in this procedure, is obtained as $ m_\Xi=1322.2$~MeV and is close to the experimental value of $1318.3$~MeV.

\begin{figure}[t]
    \centering
    \includegraphics[width=0.99\linewidth]{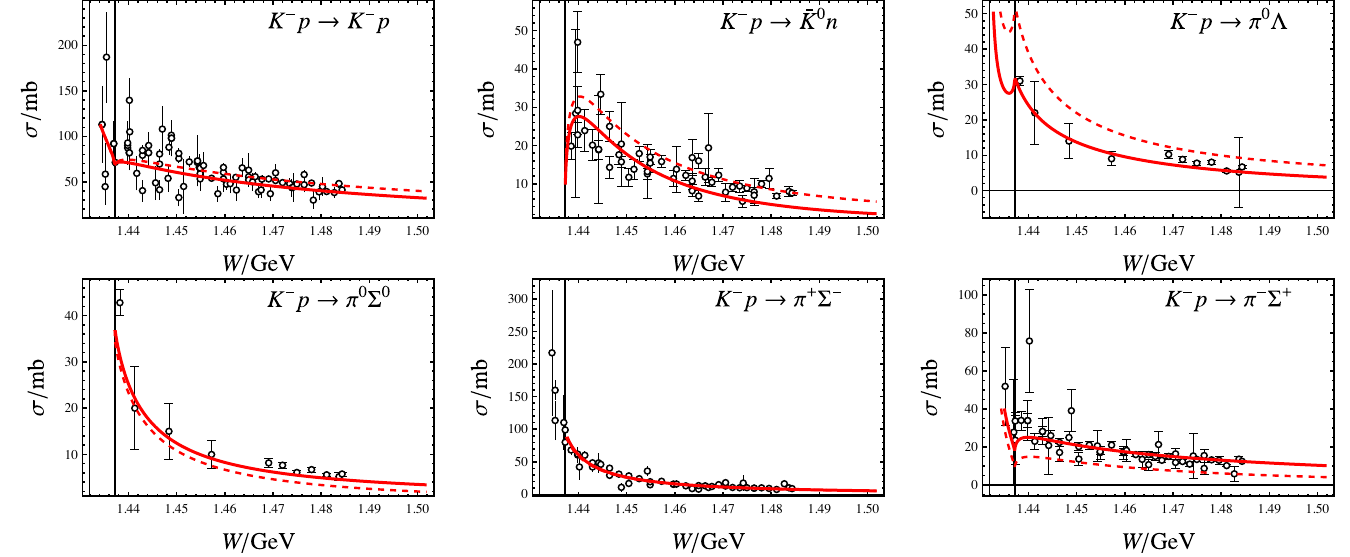}
	\caption{Fit to the experimental data for $K^-p$ scattering. For the source of the data, see the review~\cite{Mai:2020ltx}. Solid (dashed) lines: NLO (WT) fit.
	\label{fig:fit}}
\end{figure}

We fix the subtraction constant in each channel with the scale of dimensional regularization equal to the physical baryon mass
in that channel from the condition in Eq.~\eqref{eq:natural_subt},  
\begin{align}
    a_{\pi\Sigma} &= -0.70\,, \qquad a_{\bar{K}N} = -1.15\,, \qquad  a_{\eta\Lambda} = -1.21\,, \qquad ~a_{K\Xi} = -1.13\,.
\end{align}
At the SU(3) symmetric point these parameters approach a common value of 
\begin{align}
    a_{\rm SU(3)}=-0.92\,,
\end{align}
through the condition in Eq.~\eqref{eq:natural_subt}. Including the NLO part~\eqref{eq:VNLO} requires additionally the knowledge of further (dynamical and symmetry-breaking) LECs. For the symmetry breakers, we take the values determined through the running of the baryon masses, quoted in Table~\ref{tab:params_extra}. 
The dynamical LECs are determined through a fit to experimental data at the physical point, including the usual total cross sections (restricting laboratory momentum of the kaon to $P_{\rm LAB}=300$~MeV, see Fig.~\ref{fig:fit}) and the usual threshold values, including kaonic hydrogen data from Ref.~\cite{SIDDHARTA:2011dsy}. Provided the fact that Born terms are excluded from the potential kernel and the number of free parameters ($d_1,d_2,d_3,d_4$) is small, we obtained a reasonable fit for 
\begin{align}
    d_1 = -0.36\,,~~d_2=0.09\,,~~d_3 = 0.10\,,~~d_4=-0.59 .
\end{align}
in units of GeV$^{-1}$. 
We note that the obtained values are of natural size and that the minimum of the fit $\chi^2$ value is well pronounced. For the kaonic hydrogen and the threshold ratios we get $\Delta E-i\Gamma/2=356-i\,464$~eV, $\gamma=2.38$, $R_c=0.19$, $R_n=0.65$ compared to the experimental values $\Delta E-i\Gamma/2=283\pm42-i\,542\pm110$~eV, $\gamma=2.38\pm0.04$, $R_c=0.19\pm0.02$, $R_n=0.66\pm0.01$. While this fit is not perfect, we do not expect any qualitative changes of the results to be discussed.

\subsection{Pole positions}

For the physical hadron masses listed in Table~\ref{tab:mass}, the pole positions of the $\Lambda(1380)$, $\Lambda(1405)$ and $\Lambda(1680)$ resonances at LO and NLO are found as
\begin{align}
    E_{\Lambda(1380)}^{\rm LO} &= 1403.3-i\,80.3~\mathrm{MeV}, \qquad E_{\Lambda(1380)}^{\rm NLO} = 1415.4-i\,165.7~\mathrm{MeV},\nonumber\\
    E_{\Lambda(1405)}^{\rm LO} &= 1422.7-i\,16.2~\mathrm{MeV}, \qquad E_{\Lambda(1405)}^{\rm NLO} = 1417.9-i\,15.7~\mathrm{MeV}, \\
    E_{\Lambda(1680)}^{\rm LO} &= 1717.4-i\,22.9~\mathrm{MeV}, \qquad E_{\Lambda(1680)}^{\rm NLO} = 1725.9-i\,13.7~\mathrm{MeV}.\nonumber
\end{align}
The first two resonance poles are located between the $\pi\Sigma$ and $\bar{K}N$ thresholds. We note that the heavier pole agree better with the values quoted by the PDG~\cite{ParticleDataGroup:2022pth}. The width of the broad $\Lambda(1380)$ changes rapidly between the LO and NLO estimations, which is also quite uncertain in the literature values~\cite{ParticleDataGroup:2022pth}, see also the discussion in Ref.~\cite{Mai:2020ltx}. 
The situation is similar to the lower pole of the scalar charmed meson $D_0^*(2300)$: the width increases sizably from the LO calculation~\cite{Guo:2006fu} to the NLO one~\cite{Albaladejo:2016lbb}.
For the purpose of the present paper, a discussion of the SU(3) trajectories, we consider the determined pole positions as a fair representation of the realistic values.  Additionally, using the decay constants and hadron masses from Refs.~\cite{Bulava:2023rmn,Bulava:2023gfx} we have also extracted the poles using our coupled-channel chiral unitary approach, see \ref{app}, noting a number of systematic effects that can affect the positions and the nature of the these poles.

\begin{figure}[tbh]
    \centering
    \includegraphics[width=\linewidth]{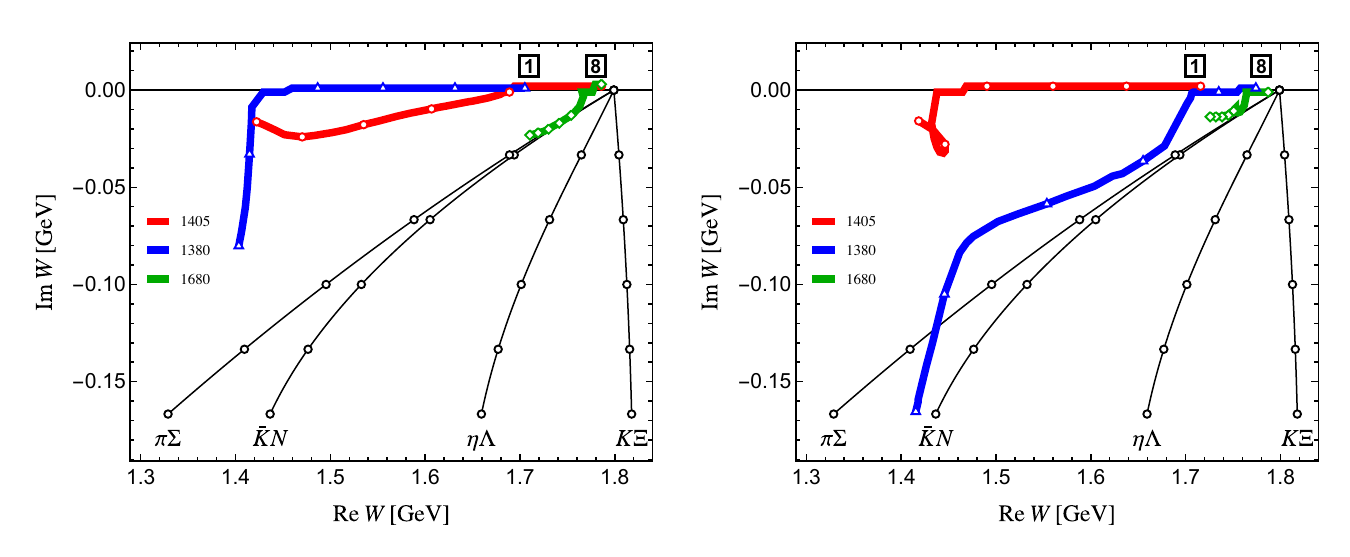}
    \caption{Motion of the poles from the SU(3) limit to the physical values of the particle masses ($(0.0,0.2,...,1.0)$ steps are shown by the empty dots). The blue, red, green lines denote the $\Lambda(1380)$, $\Lambda(1405)$ and $\Lambda(1680)$ in order. Left panel: WT-interaction. Right panel: including the NLO terms. Upper/Lower half-plane correspond to physical/unphysical sheets (small displacement along the real energy axis is added/subtracted for clarity). In both figures, numbers in black boxes denote the multiplet to which the poles belong to in the SU(3) limit. The various meson-baryon thresholds are also shown as the black solid lines.}
    \label{fig:poles trajactory}
\end{figure}

Next, we consider the SU(3) limit, noting first that there is only one two-body threshold at $E\approx 1799\,$MeV. This implies that while there are originally $2^4$ Riemann sheets, they collapse in the SU(3) limit to only two. Looking for the physically relevant pole content we find in both (LO and NLO) approaches three poles on the real axis. Notably, for the former scenario the previously discussed accidental symmetry of the  WT terms prohibits the $8/8^\prime$ mixing and leads to two degenerate octet poles. Specifically, we record three poles on the physical Riemann sheet for the LO WT-interaction: $\{E^1=1704\,{\rm MeV},E^{8}=1788\,{\rm MeV}, E^{8^{\prime}}=1788\,{\rm MeV} \}$. Including the NLO terms breaks this accidental symmetry, and we record three separated poles $\{E^1=1716\,{\rm MeV},E^{8}=1772\,{\rm MeV}, E^{8}=1787\,{\rm MeV} \}$ where both the former poles are located on the physical Riemann sheet and the latter one is located on the unphysical sheet. We note that overall the shifts due to the NLO terms are rather small in this limit.

\begin{figure}[tb]
   \centering
    \includegraphics[width=0.5\linewidth]{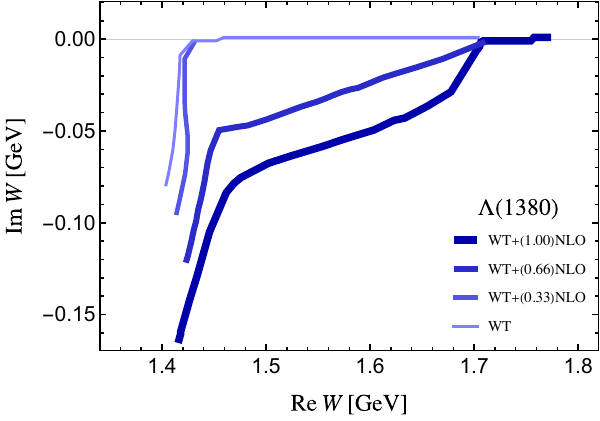}
    \caption{Trajectories of the $\Lambda(1380)$  pole from the SU(3) limit for smoothly varied NLO terms. The numbers in front of ``NLO" in the legend denote the prefactor multiplied to the NLO potential. }
    \label{fig:polemove-2}
\end{figure}

An interesting observation can be made when trying to connect the three poles in the SU(3) limit to those at the physical point. This is depicted in Fig.~\ref{fig:poles trajactory}. We observe there that the trajectory of the $\Lambda(1680)$ stays essentially the same in the LO and NLO formalisms up to the fact that the latter does not seem to reach into the physical sheet and the state remains a virtual state. Since the remaining path of the pole is quite small, we argue that this may be due to the details of the model (goodness of the NLO fit, inclusion of Born terms etc.). However,  dramatic changes due to the inclusion of the NLO terms occur for the trajectories of the $\Lambda(1380)$ and $\Lambda(1405)$. There, we see that the SU(3) limit results are not that different, but the poles in that limit extend to different poles at the physical point. 
In other words, the $\Lambda(1380)$ originates in LO/NLO from the singlet/octet state in the SU(3) limit, while $\Lambda(1405)$ originates in LO/NLO from the octet/singlet state in the SU(3) limit. This interchange of trajectories is a surprising new feature not expected from the pioneering study of Ref.~\cite{Jido:2003cb}. Obviously, one can argue that the fit is not perfect and more details or possibly next-to-next-to-leading order terms need to be included. Additionally, close to the SU(3) limit thresholds are very close to each other, such that identification of individual poles is complicated.  Still, indubitably the pole trajectories can change their behavior, which is also demonstrated in Fig.~\ref{fig:polemove-2}. There, as an example we begin with the NLO trajectory of the $\Lambda(1380)$ but scale then the NLO terms with a prefactor $x\in[0,1]$, i.e., for $x=0/1$ one obtains back the LO/NLO trajectory. We observe that between $x=0.33$ and $x=0.66$ the trajectory indeed changes from that connected to the more bound state to the less bound one in the SU(3) limit. In fact, this  interconnection between these trajectories of the $\Lambda(1380)$ and the $\Lambda(1405)$ is quite reasonable from the recent detailed study of the correlations between $\Lambda(1380)$ and $\Lambda(1405)$ in Ref.~\cite{Sadasivan:2022srs}.

\section{Summary}\label{sect:summary}

It has been suggested since long that the $\Lambda(1405)$ has a two-pole structure, thus corresponding to two states $\Lambda(1380)$ and $\Lambda(1405)$, and the lowest negative-parity baryons are dynamically generated from unitarizing the chiral amplitude.
In quark model, baryons composed of three quarks can form an SU(3) flavor singlet and two SU(3) octets.
In the dynamical generation picture, among all the SU(3) multiplets, also only the singlet and two octets have $S$-wave attractive interactions from the LO chiral dynamics and lead to dynamically generated states. 
While the mass differences among states in different multiplets depend on quark model details in the three-quark picture, they follow chiral dynamics in the dynamical generation picture.
This essential difference can be checked using lattice calculations by computing baryon spectrum in the SU(3) limit and by varying quark masses from the SU(3) limit to more realistic values,
see Refs.~\cite{Bulava:2023rmn,Bulava:2023gfx} for a recent attempt.

In this study, we have discussed the $\Lambda(1380)$, $\Lambda(1405)$ and $\Lambda(1680)$ resonances in the SU(3) limit and the evolution of their poles away from the SU(3) limit within chiral SU(3) dynamics.
The poles are generated by unitarizing the LO chiral amplitude in the WT approximation and also including the NLO terms.
We find that the $\Lambda(1405)$ state has always two different poles for any SU(3) limit, one in singlet and the other in octet, and the $\Lambda(1680)$ is degenerate with the heavier pole at LO in that limit.
The degeneracy is an artifect of unitarizing only the WT  chiral amplitude at LO and is  removed by considering higher order contributions (even by considering the Born
terms at LO).
We have worked out the pole trajectories corresponding to these states when one moves away from the SU(3) limit.  Most interestingly, we find that the trajectories of the $\Lambda(1380)$ and the $\Lambda(1405)$ change roles when
going from the WT term at LO to the NLO contributions. This phenomenon
can be further studied on the lattice when starting with an SU(3) symmetric
configuration and then evolving to the physical light and strange quark masses.

\section*{Acknowledgements}

We thank Tetsuo Hyodo for contributions during the early stage of this work, which started in 2019. We also thank Ale$\check{\rm s}$ Ciepl\'y and Peter Bruns for useful discussions on related topics. This work is supported in
part by  the DFG (Project number 196253076 - TRR 110)
and the NSFC (Grant No. 11621131001) through the funds provided
to the Sino-German CRC 110 ``Symmetries and the Emergence of
Structure in QCD",  by the Chinese Academy of Sciences (CAS) through a President's International Fellowship Initiative (PIFI) (Grant No. 2018DM0034), by the VolkswagenStiftung (Grant No. 93562), by the EU Horizon 2020 research and innovation programme, STRONG-2020 project
under grant agreement No. 824093, by the NSFC under Grant Nos.~12125507, 11835015 and 12047503, and by CAS under Grant Nos. YSBR-101 and XDB34030000.

\begin{appendix}

\section{Investigating the Baryon Scattering Collaboration poles}
\label{app}
The Baryon Scattering Collaboration~\cite{Bulava:2023rmn,Bulava:2023gfx} has performed a two-channel analysis ($\pi\Sigma$-$\bar KN$) in the region of the $\Lambda(1405)$ for unphysical hadron masses and decay constants $M_\pi=203.7$~MeV, $M_K=486.4$~MeV, $M_\eta=551.1$~MeV, $F_\pi=93.2$~MeV, $F_K=108.2$~MeV, $m_N=979.8$~MeV, $m_\Lambda=1132.8$~MeV, $m_\Sigma=1193.9$~MeV. For the remaining required parameters we take the usual values $F_\eta=1.3F_\pi$, $M_\Xi=1322.2$~MeV. With these values we solve the Bethe-Salpeter equation~\eqref{eq:BSE} at leading order (WT-interaction) and including the NLO interactions. 

\begin{figure}[htb]
    \centering
    \includegraphics[width=0.80\linewidth]{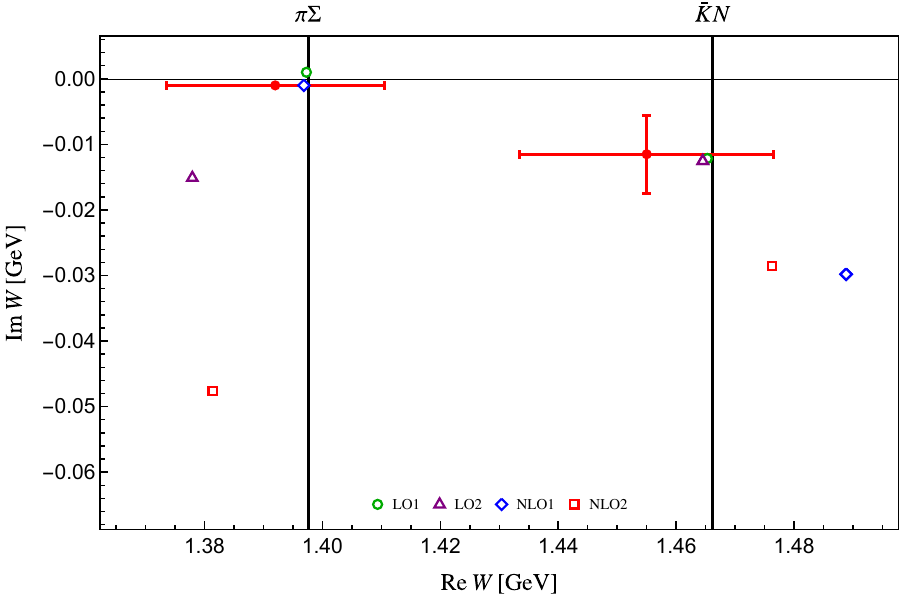}
    \caption{Predictions for the poles of the $\Lambda(1380)$ and $\Lambda(1405)$ for the lattice setup of Refs.~\cite{Bulava:2023rmn,Bulava:2023gfx}. Green circles/blue diamonds: WT/NLO-interaction with the subtraction constants at the physical point. Purple triangles/red squares: WT/NLO-interaction with the subtraction constants using Eq.~\eqref{eq:natural_subt} with $\mu$ set to the corresponding lattice baryon mass.   
    Results of Ref.~\cite{Bulava:2023rmn} are depicted by the red dots with statistical and systematic error bars added in quadrature. Upper/lower half plane correspond to the physical/unphysical $\{-+++\}$ Riemann sheets. Small displacements along the real axis are added for clarity. }
    \label{fig:CLS}
\end{figure}

In exploring lattice setups with chiral unitary approaches certain subtleties can arise in, e.g., how the decay constants are used or how the lattice errors on the hadron masses are propagated, see Ref.~\cite{Mai:2019pqr} for an in-depth discussion of such effects. In the current work we have explored two ways to give a sense of the systematic uncertainty. Namely, we use either subtraction constants at the physical point or those from the condition in Eq.~\eqref{eq:natural_subt} for the WT and NLO-interactions, correspondingly. This yields the pole positions on the physical (top half-plane) and unphysical ($\{-+++\}$ lower half-plane) Riemann sheet as depicted in Fig.~\ref{fig:CLS}. We note that in certain scenarios the high-mass or the low-mass pole agrees astonishingly well with the analysis of Ref.~\cite{Bulava:2023rmn}. Still it is important to note that through a chiral trajectory poles can transition from the unphysical Riemann sheet to the physical, c.f. NLO1 vs. LO1 solutions for the $\Lambda(1380)$ in Fig.~\ref{fig:CLS}. From the findings displayed here it is obvious that one cannot make a definite statement about the nature of the low-mass pole. In the future one should reanalyze the energy levels directly using a coupled-channel chiral approach constrained by experimental data.

\end{appendix}

\bibliographystyle{elsarticle-num}
\bibliography{BIB}

\begin{thebibliography}{10}
\expandafter\ifx\csname url\endcsname\relax
  \def\url#1{\texttt{#1}}\fi
\expandafter\ifx\csname urlprefix\endcsname\relax\def\urlprefix{URL }\fi
\expandafter\ifx\csname href\endcsname\relax
  \def\href#1#2{#2} \def\path#1{#1}\fi

\bibitem{Oller:2000fj}
J.~A. Oller, U.-G. Mei{\ss}ner, {Chiral dynamics in the presence of bound
  states: Kaon nucleon interactions revisited}, Phys. Lett. B 500 (2001)
  263--272.
\newblock \href {http://arxiv.org/abs/hep-ph/0011146}
  {\path{arXiv:hep-ph/0011146}}, \href
  {https://doi.org/10.1016/S0370-2693(01)00078-8}
  {\path{doi:10.1016/S0370-2693(01)00078-8}}.

\bibitem{Jido:2003cb}
D.~Jido, J.~A. Oller, E.~Oset, A.~Ramos, U.-G. Mei{\ss}ner, {Chiral dynamics of
  the two $\Lambda(1405)$ states}, Nucl. Phys. A 725 (2003) 181--200.
\newblock \href {http://arxiv.org/abs/nucl-th/0303062}
  {\path{arXiv:nucl-th/0303062}}, \href
  {https://doi.org/10.1016/S0375-9474(03)01598-7}
  {\path{doi:10.1016/S0375-9474(03)01598-7}}.

\bibitem{ParticleDataGroup:2022pth}
R.~L. Workman, et~al., {Review of Particle Physics}, PTEP 2022 (2022) 083C01.
\newblock \href {https://doi.org/10.1093/ptep/ptac097}
  {\path{doi:10.1093/ptep/ptac097}}.

\bibitem{Mai:2020ltx}
M.~Mai, {Review of the ${\Lambda }$(1405) A curious case of a strangeness
  resonance}, Eur. Phys. J. ST 230~(6) (2021) 1593--1607.
\newblock \href {http://arxiv.org/abs/2010.00056} {\path{arXiv:2010.00056}},
  \href {https://doi.org/10.1140/epjs/s11734-021-00144-7}
  {\path{doi:10.1140/epjs/s11734-021-00144-7}}.

\bibitem{Hyodo:2020czb}
T.~Hyodo, M.~Niiyama, {QCD and the strange baryon spectrum}, Prog. Part. Nucl.
  Phys. 120 (2021) 103868.
\newblock \href {http://arxiv.org/abs/2010.07592} {\path{arXiv:2010.07592}},
  \href {https://doi.org/10.1016/j.ppnp.2021.103868}
  {\path{doi:10.1016/j.ppnp.2021.103868}}.

\bibitem{Meissner:2020khl}
U.-G. Mei\ss{}ner, {Two-pole structures in QCD: Facts, not fantasy!}, Symmetry
  12~(6) (2020) 981.
\newblock \href {http://arxiv.org/abs/2005.06909} {\path{arXiv:2005.06909}},
  \href {https://doi.org/10.3390/sym12060981} {\path{doi:10.3390/sym12060981}}.

\bibitem{Lu:2022hwm}
J.-X. Lu, L.-S. Geng, M.~D{\"o}ring, M.~Mai, {Cross-Channel Constraints on
  Resonant Antikaon-Nucleon Scattering}, Phys. Rev. Lett. 130~(7) (2023)
  071902.
\newblock \href {http://arxiv.org/abs/2209.02471} {\path{arXiv:2209.02471}},
  \href {https://doi.org/10.1103/PhysRevLett.130.071902}
  {\path{doi:10.1103/PhysRevLett.130.071902}}.

\bibitem{Sadasivan:2022srs}
D.~Sadasivan, M.~Mai, M.~D\"oring, U.-G. Mei\ss{}ner, F.~Amorim, J.~P. Klucik,
  J.-X. Lu, L.-S. Geng, {New insights into the pole parameters of the
  $\Lambda(1380)$, the $\Lambda(1405)$ and the $\Sigma(1385)$}, Front. Phys. 11
  (2023) 1139236.
\newblock \href {http://arxiv.org/abs/2212.10415} {\path{arXiv:2212.10415}},
  \href {https://doi.org/10.3389/fphy.2023.1139236}
  {\path{doi:10.3389/fphy.2023.1139236}}.

\bibitem{Engel:2012qp}
G.~P. Engel, C.~B. Lang, A.~Sch\"afer, {Low-lying $\Lambda$ baryons from the
  lattice}, Phys. Rev. D 87~(3) (2013) 034502.
\newblock \href {http://arxiv.org/abs/1212.2032} {\path{arXiv:1212.2032}},
  \href {https://doi.org/10.1103/PhysRevD.87.034502}
  {\path{doi:10.1103/PhysRevD.87.034502}}.

\bibitem{Hall:2014uca}
J.~M.~M. Hall, W.~Kamleh, D.~B. Leinweber, B.~J. Menadue, B.~J. Owen, A.~W.
  Thomas, R.~D. Young, {Lattice QCD Evidence that the
  \ensuremath{\Lambda}(1405) Resonance is an Antikaon-Nucleon Molecule}, Phys.
  Rev. Lett. 114~(13) (2015) 132002.
\newblock \href {http://arxiv.org/abs/1411.3402} {\path{arXiv:1411.3402}},
  \href {https://doi.org/10.1103/PhysRevLett.114.132002}
  {\path{doi:10.1103/PhysRevLett.114.132002}}.

\bibitem{Bulava:2023rmn}
J.~Bulava, et~al., {The two-pole nature of the $\Lambda(1405)$ from lattice
  QCD} (7 2023).
\newblock \href {http://arxiv.org/abs/2307.10413} {\path{arXiv:2307.10413}}.

\bibitem{Bulava:2023gfx}
J.~Bulava, et~al., {Lattice QCD study of $\pi\Sigma-\bar{K}N$ scattering and
  the $\Lambda(1405)$ resonance} (7 2023).
\newblock \href {http://arxiv.org/abs/2307.13471} {\path{arXiv:2307.13471}}.

\bibitem{Bietenholz:2011qq}
W.~Bietenholz, et~al., {Flavour blindness and patterns of flavour symmetry
  breaking in lattice simulations of up, down and strange quarks}, Phys. Rev. D
  84 (2011) 054509.
\newblock \href {http://arxiv.org/abs/1102.5300} {\path{arXiv:1102.5300}},
  \href {https://doi.org/10.1103/PhysRevD.84.054509}
  {\path{doi:10.1103/PhysRevD.84.054509}}.

\bibitem{Nagahiro:2011jn}
H.~Nagahiro, K.~Nawa, S.~Ozaki, D.~Jido, A.~Hosaka, {Composite and elementary
  natures of $a_1(1260)$ meson}, Phys. Rev. D 83 (2011) 111504.
\newblock \href {http://arxiv.org/abs/1101.3623} {\path{arXiv:1101.3623}},
  \href {https://doi.org/10.1103/PhysRevD.83.111504}
  {\path{doi:10.1103/PhysRevD.83.111504}}.

\bibitem{Garcia-Recio:2003ejq}
C.~Garcia-Recio, M.~F.~M. Lutz, J.~Nieves, {Quark mass dependence of s wave
  baryon resonances}, Phys. Lett. B 582 (2004) 49--54.
\newblock \href {http://arxiv.org/abs/nucl-th/0305100}
  {\path{arXiv:nucl-th/0305100}}, \href
  {https://doi.org/10.1016/j.physletb.2003.11.073}
  {\path{doi:10.1016/j.physletb.2003.11.073}}.

\bibitem{Molina:2015uqp}
R.~Molina, M.~D\"oring, {Pole structure of the $\Lambda$(1405) in a recent QCD
  simulation}, Phys. Rev. D 94~(5) (2016) 056010, [Addendum: Phys.Rev.D 94,
  079901 (2016)].
\newblock \href {http://arxiv.org/abs/1512.05831} {\path{arXiv:1512.05831}},
  \href {https://doi.org/10.1103/PhysRevD.94.079901}
  {\path{doi:10.1103/PhysRevD.94.079901}}.

\bibitem{Bruns:2021krp}
P.~C. Bruns, A.~Ciepl\'y, {SU(3) flavor symmetry considerations for the
  K\textasciimacron{}N coupled channels system}, Nucl. Phys. A 1019 (2022)
  122378.
\newblock \href {http://arxiv.org/abs/2109.03109} {\path{arXiv:2109.03109}},
  \href {https://doi.org/10.1016/j.nuclphysa.2021.122378}
  {\path{doi:10.1016/j.nuclphysa.2021.122378}}.

\bibitem{Xie:2023cej}
J.-M. Xie, J.-X. Lu, L.-S. Geng, B.-S. Zou, {Two-pole structures as a universal
  phenomenon dictated by coupled-channel chiral dynamics} (7 2023).
\newblock \href {http://arxiv.org/abs/2307.11631} {\path{arXiv:2307.11631}}.

\bibitem{Oset:1997it}
E.~Oset, A.~Ramos, {Nonperturbative chiral approach to $S$-wave $\bar KN$
  interactions}, Nucl. Phys. A 635 (1998) 99--120.
\newblock \href {http://arxiv.org/abs/nucl-th/9711022}
  {\path{arXiv:nucl-th/9711022}}, \href
  {https://doi.org/10.1016/S0375-9474(98)00170-5}
  {\path{doi:10.1016/S0375-9474(98)00170-5}}.

\bibitem{Borasoy:2005ie}
B.~Borasoy, R.~Nissler, W.~Weise, {Chiral dynamics of kaon-nucleon
  interactions, revisited}, Eur. Phys. J. A 25 (2005) 79--96.
\newblock \href {http://arxiv.org/abs/hep-ph/0505239}
  {\path{arXiv:hep-ph/0505239}}, \href
  {https://doi.org/10.1140/epja/i2005-10079-1}
  {\path{doi:10.1140/epja/i2005-10079-1}}.

\bibitem{Hyodo:2006kg}
T.~Hyodo, D.~Jido, A.~Hosaka, {Study of exotic hadrons in s-wave scatterings
  induced by chiral interaction in the flavor symmetric limit}, Phys. Rev. D 75
  (2007) 034002.
\newblock \href {http://arxiv.org/abs/hep-ph/0611004}
  {\path{arXiv:hep-ph/0611004}}, \href
  {https://doi.org/10.1103/PhysRevD.75.034002}
  {\path{doi:10.1103/PhysRevD.75.034002}}.

\bibitem{Roca:2008kr}
L.~Roca, T.~Hyodo, D.~Jido, {On the nature of the $\Lambda(1405)$ and
  $\Lambda(1670)$ from their $N_c$ behavior in chiral dynamics}, Nucl. Phys. A
  809 (2008) 65--87.
\newblock \href {http://arxiv.org/abs/0804.1210} {\path{arXiv:0804.1210}},
  \href {https://doi.org/10.1016/j.nuclphysa.2008.05.014}
  {\path{doi:10.1016/j.nuclphysa.2008.05.014}}.

\bibitem{Bruns:2010sv}
P.~C. Bruns, M.~Mai, U.-G. Mei{\ss}ner, {Chiral dynamics of the S11(1535) and
  S11(1650) resonances revisited}, Phys. Lett. B 697 (2011) 254--259.
\newblock \href {http://arxiv.org/abs/1012.2233} {\path{arXiv:1012.2233}},
  \href {https://doi.org/10.1016/j.physletb.2011.02.008}
  {\path{doi:10.1016/j.physletb.2011.02.008}}.

\bibitem{Mai:2012dt}
M.~Mai, U.-G. Mei{\ss}ner, {New insights into antikaon-nucleon scattering and
  the structure of the Lambda(1405)}, Nucl. Phys. A 900 (2013) 51 -- 64.
\newblock \href {http://arxiv.org/abs/1202.2030} {\path{arXiv:1202.2030}},
  \href {https://doi.org/10.1016/j.nuclphysa.2013.01.032}
  {\path{doi:10.1016/j.nuclphysa.2013.01.032}}.

\bibitem{Sadasivan:2018jig}
D.~Sadasivan, M.~Mai, M.~D\"oring, {S- and p-wave structure of $S=-1$
  meson-baryon scattering in the resonance region}, Phys. Lett. B 789 (2019)
  329--335.
\newblock \href {http://arxiv.org/abs/1805.04534} {\path{arXiv:1805.04534}},
  \href {https://doi.org/10.1016/j.physletb.2018.12.035}
  {\path{doi:10.1016/j.physletb.2018.12.035}}.

\bibitem{Lutz:2001yb}
M.~F.~M. Lutz, E.~E. Kolomeitsev, {Relativistic chiral SU(3) symmetry, large
  $N_c$ sum rules and meson baryon scattering}, Nucl. Phys. A 700 (2002)
  193--308.
\newblock \href {http://arxiv.org/abs/nucl-th/0105042}
  {\path{arXiv:nucl-th/0105042}}, \href
  {https://doi.org/10.1016/S0375-9474(01)01312-4}
  {\path{doi:10.1016/S0375-9474(01)01312-4}}.

\bibitem{Hyodo:2008xr}
T.~Hyodo, D.~Jido, A.~Hosaka, {Origin of the resonances in the chiral unitary
  approach}, Phys. Rev. C 78 (2008) 025203.
\newblock \href {http://arxiv.org/abs/0803.2550} {\path{arXiv:0803.2550}},
  \href {https://doi.org/10.1103/PhysRevC.78.025203}
  {\path{doi:10.1103/PhysRevC.78.025203}}.

\bibitem{Gasser:1984gg}
J.~Gasser, H.~Leutwyler, {Chiral Perturbation Theory: Expansions in the Mass of
  the Strange Quark}, Nucl. Phys. B 250 (1985) 465--516.
\newblock \href {https://doi.org/10.1016/0550-3213(85)90492-4}
  {\path{doi:10.1016/0550-3213(85)90492-4}}.

\bibitem{Nebreda:2010wv}
J.~Nebreda, J.~R. Pel{\'a}ez, {Strange and non-strange quark mass dependence of
  elastic light resonances from SU(3) Unitarized Chiral Perturbation Theory to
  one loop}, Phys. Rev. D 81 (2010) 054035.
\newblock \href {http://arxiv.org/abs/1001.5237} {\path{arXiv:1001.5237}},
  \href {https://doi.org/10.1103/PhysRevD.81.054035}
  {\path{doi:10.1103/PhysRevD.81.054035}}.

\bibitem{Borasoy:1996bx}
B.~Borasoy, U.-G. Mei{\ss}ner, {Chiral Expansion of Baryon Masses and
  \ensuremath{\sigma}-Terms}, Annals Phys. 254 (1997) 192--232.
\newblock \href {http://arxiv.org/abs/hep-ph/9607432}
  {\path{arXiv:hep-ph/9607432}}, \href {https://doi.org/10.1006/aphy.1996.5630}
  {\path{doi:10.1006/aphy.1996.5630}}.

\bibitem{Frink:2004ic}
M.~Frink, U.-G. Mei{\ss}ner, {Chiral extrapolations of baryon masses for
  unquenched three flavor lattice simulations}, JHEP 07 (2004) 028.
\newblock \href {http://arxiv.org/abs/hep-lat/0404018}
  {\path{arXiv:hep-lat/0404018}}, \href
  {https://doi.org/10.1088/1126-6708/2004/07/028}
  {\path{doi:10.1088/1126-6708/2004/07/028}}.

\bibitem{Hoferichter:2015hva}
M.~Hoferichter, J.~Ruiz~de Elvira, B.~Kubis, U.-G. Mei\ss{}ner,
  {Roy\textendash{}Steiner-equation analysis of pion\textendash{}nucleon
  scattering}, Phys. Rept. 625 (2016) 1--88.
\newblock \href {http://arxiv.org/abs/1510.06039} {\path{arXiv:1510.06039}},
  \href {https://doi.org/10.1016/j.physrep.2016.02.002}
  {\path{doi:10.1016/j.physrep.2016.02.002}}.

\bibitem{SIDDHARTA:2011dsy}
M.~Bazzi, et~al., {A New Measurement of Kaonic Hydrogen X-rays}, Phys. Lett. B
  704 (2011) 113--117.
\newblock \href {http://arxiv.org/abs/1105.3090} {\path{arXiv:1105.3090}},
  \href {https://doi.org/10.1016/j.physletb.2011.09.011}
  {\path{doi:10.1016/j.physletb.2011.09.011}}.

\bibitem{Guo:2006fu}
F.-K. Guo, P.-N. Shen, H.-C. Chiang, R.-G. Ping, B.-S. Zou, {Dynamically
  generated $0^+$ heavy mesons in a heavy chiral unitary approach}, Phys. Lett.
  B 641 (2006) 278--285.
\newblock \href {http://arxiv.org/abs/hep-ph/0603072}
  {\path{arXiv:hep-ph/0603072}}, \href
  {https://doi.org/10.1016/j.physletb.2006.08.064}
  {\path{doi:10.1016/j.physletb.2006.08.064}}.

\bibitem{Albaladejo:2016lbb}
M.~Albaladejo, P.~Fernandez-Soler, F.-K. Guo, J.~Nieves, {Two-pole structure of
  the $D^\ast_0(2400)$}, Phys. Lett. B 767 (2017) 465--469.
\newblock \href {http://arxiv.org/abs/1610.06727} {\path{arXiv:1610.06727}},
  \href {https://doi.org/10.1016/j.physletb.2017.02.036}
  {\path{doi:10.1016/j.physletb.2017.02.036}}.

\bibitem{Mai:2019pqr}
M.~Mai, C.~Culver, A.~Alexandru, M.~D\"oring, F.~X. Lee, {Cross-channel study
  of pion scattering from lattice QCD}, Phys. Rev. D 100~(11) (2019) 114514.
\newblock \href {http://arxiv.org/abs/1908.01847} {\path{arXiv:1908.01847}},
  \href {https://doi.org/10.1103/PhysRevD.100.114514}
  {\path{doi:10.1103/PhysRevD.100.114514}}.

\end{thebibliography}

\end{document}